\documentclass[11pt,a4paper]{scrartcl}

\usepackage{ILD}

\usepackage[symbol]{footmisc} 
\usepackage{feynmf}
\usepackage{textpos}
\usepackage{lineno}
\usepackage{hyperref}
\usepackage{dirtytalk}

\title{CP violation in the Higgs sector at ILC}

\ildproc{phys}{2021}{011}

\date{\today}

\addauthor{T. Agatonovi\'{c}-Jovin}{\institute{1}}
\addauthor{N. Vuka\u{s}inovi\'{c}}{\thanks{nvukasinovic@vin.bg.ac.rs}\institute{1}}
\addauthor{I. Bo\v{z}ovi\'{c}-Jelisav\u{c}i\'{c}}{\institute{1}}
\addauthor{G. Ka\u{c}arevi\'{c}}{\institute{1}}
\addauthor{M. Radulovi\'{c}}{\institute{2}}
\addauthor{J. Stevanovi\'{c}}{\institute{2}}
\addauthor{G. Milutinovi\'{c}-Dumbelovi\'{c}}{\institute{1}}
\addauthor{I. Smiljani\'{c}}{\institute{1}}

\addinstitute{1}{\say{VIN\u{C}A} Institute of Nuclear Sciences - National Institute of the Republic of Serbia, University of Belgrade\\
  Mike Petrovi\'{c}a Alasa 12-14, Belgrade, Serbia}
  \addinstitute{2}{University of Kragujevac, Faculty of Science\\
Radoja Domanovi\'{c}a 12, Kragujevac, Serbia}

\onbehalfof{the ILD concept group and ILC/IDT physics and detector group}

\abstract{ CP violation is one of Sakharov's conditions for the matter-antimatter asymmetry of the Universe. The experimentally observed size of CP violation is insufficient to account for this. Is CP violated in the Higgs sector? Could the SM-like Higgs boson be a mixture of CP even and CP odd states of an extended Higgs sector? With what precision could such effects be measured at future electron-positron colliders? These questions will be discussed in the light of the latest studies at ILC. }

\titlecomment{Poster presented at the European Physical Society conference on high energy physics (EPS-HEP2021), Online conference, July 26-30, 2021.}

\graphicspath{ {./logos/}{./figures/} }

\begin{document}

\titlepage  

\section{CP Violation}

Experimentally observed size of CP violation (CPV) is not supported by the formalism of the Standard Model (SM). New sources of CPV
beyond the SM are necessary in order to explain the still open issue of baryon asymmetry
of the Universe.
CPV in the Higgs sector can be probed in the interaction of Higgs boson with vector bosons or
fermions. The SM Lagrangian can be modified by adding CP violating terms
at a loop level in case of \emph{HVV} vertices or at the Born level in case of \emph{Hff} vertices.

So far, there are only few results on measurement of the CPV mixing angle ($\Psi_{\mathrm{CP}}$) between the Higgs scalar and pseudoscalar states.
Among the future projects (HL-LHC, HE-LHC, CEPC, FCC-ee$_{240}$), ILC has the most promising projection of precision of the $\Psi_{\mathrm{CP}}$ measurement, ($\Delta\Psi_{\mathrm{CP}}$ = 4$^{\circ}$), in the fermionic H $\xrightarrow{}{\tau^{+}\tau^{-}}$ decay \cite{r1}.

\section{ILC Project}

The International Linear Collider (ILC) is a high-luminosity linear e$^-$e$^+$ collider with centre-of-mass-energy
range of 250-500 GeV (extendable to 1 TeV) aimed for precision studies in the Higgs sector, operating as
a Higgs factory. The project is also optimised for top-physics and EW studies, as well as to probe new physics phenomena in a direct or indirect way. The electron beam will be polarized to 80\%, and the baseline
plan includes an undulator-based positron source which will deliver 30\% positron polarization. Realization of the ILC as a Higgs factory goes inline with the 2020 Update of the European Particle
Physics Strategy \cite{r2}.

Two detector concepts have been developed, ILD and SiD \cite{r3}, as general-purpose detectors designed to optimally address the ILC physics goals. The Particle-Flow
technique \cite{r4} that will play a central role in event reconstruction, requiring highly granular calorimeters and excellent low material budget tracking and vertexing systems in the solenoid magnetic field up to 5 T. 
 
\section{Probing CP Violation at ILC}

\begin{figure}[h]
\centering
\includegraphics[width=.4\textwidth, height=.3\textwidth]{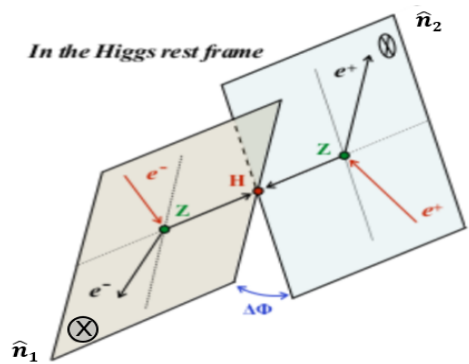}
\caption{Definition of the CPV sensitive angle $\Phi$ in the Higgs boson production in ZZ-fusion.}
\label{fig1}
\end{figure}

There are numerous Higgs production processes available at the ILC (like HZ, WW-fusion, ZZ-fusion) at
various centre-of-mass-energies, that offer a plethora of possibilities for individual measurements and
combinations of Higgs boson production and decay channels. The correlation between spin orientations of vector bosons from a Higgs production (or decay) can be extracted from the angle $\Phi$ between the production (decay) planes. Illustration of $\Phi$ definition in the Higgs production in ZZ-fusion is given in Fig. \ref{fig1}. For the same orientated unit vectors orthogonal to the production planes (as in Fig. \ref{fig1}), CP-sensitive angle between the planes can be defined as:
\begin{equation}
\label{fi}
\Phi = a \arccos(\hat{n}_{1}\cdot\hat{n}_{2})
\end{equation}
\noindent where the unit vectors are:
\begin{equation}
\label{n1}
\hat{n}_{1} = \frac{ q_{e^-_{i}}\times q_{e^-_{f}} }{ |q_{e^-_{i}}\times q_{e^-_{f}}| } \hspace{1cm} \mathrm{and} \hspace{1cm} \hat{n}_{2} = \frac{ q_{e^+_{i}}\times q_{e^+_{f}} }{ |q_{e^+_{i}}\times q_{e^+_{f}}|},
\end{equation}
\noindent and $q_{e^{-(+)}_{i}}, q_{e^{-(+)}_{f}}$ stands for momenta of initial and final e$^\mp$ states. The coefficient $a$ defines how the second (anti-fermion or an off-shell boson in case of a Higgs decay to VV$^*$) plane is rotated w.r.t. the first (fermion or an on-shell boson) plane and is defined as:
\begin{equation}
\label{a}
a = \frac{ q_{Z_{e^-}}\cdot (\hat{n}_{1} \times \hat{n}_{2})}{ |q_{Z_{e^-}}\cdot (\hat{n}_{1} \times \hat{n}_{2})|} 
\end{equation}
where $q_{Z_{e^-}}$ is momentum of the Z boson in the first plane. The direction of the Z boson momentum in the first plane (or an on-shell boson in case of a Higgs decay to VV$^*$) regulates the notion of forward and backward, where if the second plane falls backwards (as illustrated in Fig. \ref{fig1}) $a$ = $-1$, otherwise $a$ = 1.

\subsection{CP violation in $H\xrightarrow{}{\tau^{+}\tau^{-}}$ at 250 GeV}

\begin{figure}[h]
\hspace{10 mm}
\includegraphics[width=.3\textwidth, height=.25\textwidth ]{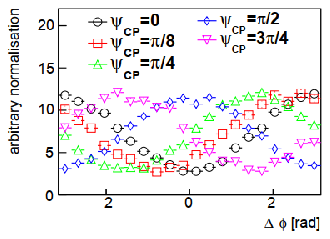} 
\hspace{3 cm}
\includegraphics[width=.3\textwidth, height=.25\textwidth]{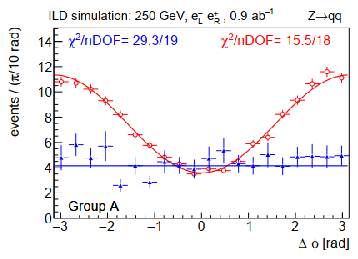}
\caption{Left: Dependence of the generated angle $\Phi$ on various assumptions on the $\Psi_{\mathrm{CP}}$ value. Right: Reconstructed $\Phi$ distribution of signal (red, fitted with a solid line) and background (blue, fitted with a flat line) events. Distributions are normalised to 0.9 ab$^{-1}$ of data in the (80, 30)\% (e$_\mathrm{L}^-$e$_\mathrm{R}^+$) beam polarisation, at 250 GeV ILC.}
\label{fig5}
\end{figure}

Higgs fermionic decay to $\tau^{+}\tau^{-}$ at 250 GeV centre-of-mass-energy provides excellent sensitivity to $\Psi_{\mathrm{CP}}$ to be extracted from the fit of CP-sensitive angle $\Phi$. Dependence of the $\Phi$ distribution on various assumptions on the CP-mixing strength ($\Psi_{\mathrm{CP}}$ value) is illustrated in Fig. \ref{fig5} (left) \cite{r5} . Figure \ref{fig5} (right) is showing reconstructed $\Phi$ of signal and background, where background $\Phi$ distribution is CP insensitive (i.e. flat). In the full simulation of ILD detector response, with 0.9 ab$^{-1}$ of data, absolute statistical uncertainty of the $\Psi_{\mathrm{CP}}$ measurement of 4$^{\circ}$ is found. There is an ongoing analysis in the same channel, at 250 GeV centre-of-mass energy with the SiD detector model \cite{r6}. 

\subsection{CP violation in HZZ production at 1 TeV}
\begin{figure}[h] 
\hspace{10 mm}
\includegraphics[width=.3\textwidth, height=.25\textwidth ]{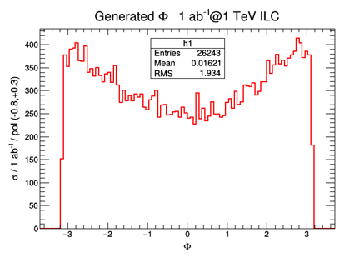}
\hspace{3 cm}
\includegraphics[width=.3\textwidth, height=.25\textwidth ]{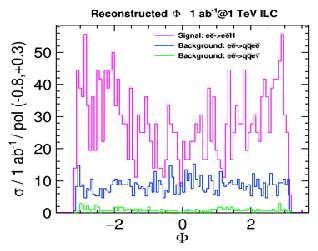}
\caption{Left: Distribution of the angle $\Phi$ at the generator level in the full physical range. Right: Reconstructed distributions of $\Phi$ in the tracker acceptance region for signal (purple) and dominant background (blue and green) processes.}
\label{fig4}
\end{figure}
There is an ongoing analysis aiming to estimate achievable precision on $\Psi_{\mathrm{CP}}$ (for $\Psi_{\mathrm{CP}}$ = 0) by reconstructing $\Phi$ between the Higgs production planes (as illustrated in Fig. \ref{fig1}), in events in which the Higgs boson is produced by ZZ-fusion at 1 TeV ILC. Full simulation of ILD detector for ILC is assumed. Figure \ref{fig4} (left) illustrates generated information on angle $\Phi$, in the full physical range of polar angles, expected with 1 ab$^{-1}$ of integrated luminosity and ($-80, +30$)\% (e$_\mathrm{L}^-$e$_\mathrm{R}^+$) beam polarization;
On Fig. \ref{fig4} (right), the reconstructed CP violating observable $\Phi$ is shown, in the tracker acceptance region\footnotemark, for the exclusive $H\xrightarrow{}b\bar{b}$ decay channel, against dominant backgrounds. Events are preselected with $\sim$ 80\% efficiency, to suppress high-cross section background processes like $e^-e^+\rightarrow qqe\nu$. As expected, background distributions are flat exhibiting no CPV sensitive structure.
\footnotetext{Tracker acceptance region assumes that both electron and positron from  $e^-e^+\rightarrow He^-e^+$ final state are reconstructed within the tracking detector, in the range of polar angles between 8$^{\circ}$ and 172$^{\circ}$ (barrel).}



\end{document}